\documentclass{article}
\usepackage[T1]{fontenc}
\usepackage[utf8]{inputenc}
\usepackage{ismir} % Remove the "submission" option for camera-ready version
\usepackage{amsmath,cite,url}
\usepackage{graphicx}
\usepackage{color}
\usepackage{tikz}
\usepackage{amssymb}
\usepackage{xcolor}
\usepackage{booktabs}  
\usepackage{multirow}   
\usepackage{array}      
\usepackage{bbm}

\usepackage[protrusion=true, expansion=true, tracking=smallcaps]{microtype}
\newcommand{\textttalt}[1]{{\fontfamily{zi4}\selectfont #1}}

\title{Scaling Self-Supervised Representation Learning\\for Symbolic Piano Performance}

\oneauthor
  {Anonymous Authors}
  {Anonymous Affiliations\\\texttt{anonymous@ismir.net}}

\multauthor
  {Louis Bradshaw$^{1,4}$ \hspace{0.3cm} Honglu Fan$^{3,4}$ \hspace{0.3cm} Alexander Spangher$^{2,4}$ \hspace{0.3cm} Stella Biderman$^4$ \hspace{0.3cm} Simon Colton$^1$}
  {
  $^1$ Queen Mary University of London \hspace{0.3cm}
  $^2$ University of Southern California\\
  $^3$ University of Geneva \hspace{0.3cm}
  $^4$ EleutherAI\\
  {\tt\small l.b.bradshaw@qmul.ac.uk, honglu.fan@unige.ch, spangher@usc.edu}
  }

\def\authorname{L. Bradshaw, H. Fan, A. Spangher, S. Biderman and S. Colton}

\usepackage[bookmarks=false,pdfauthor={\authorname},pdfsubject={\pdfsubject},hidelinks]{hyperref}
\usepackage{cleveref}
\begin{document}

\maketitle

\begin{abstract}
We study the capabilities of generative autoregressive transformer models trained on large amounts of symbolic solo-piano transcriptions. After first pretraining on approximately 60,000 hours of music, we use a comparatively smaller, high-quality subset, to finetune models to produce musical continuations, perform symbolic classification tasks, and produce general-purpose contrastive MIDI embeddings by adapting the SimCLR framework to symbolic music. When evaluating piano continuation coherence, our generative model outperforms leading symbolic generation techniques and remains competitive with proprietary audio generation models. On MIR classification benchmarks, frozen representations from our contrastive model achieve state-of-the-art results in linear probe experiments, while direct finetuning demonstrates the generalizability of pretrained representations, often requiring only a few hundred labeled examples to specialize to downstream tasks. 
\end{abstract}

\section{Introduction}

Modern machine learning systems increasingly utilize \textit{self-supervised learning} (SSL) as a core component of their training pipeline. In this paradigm, general-purpose representations are learned during an initial phase of self-guided learning, which can then be adapted to specialized tasks, often outperforming purely supervised approaches, particularly when access to supervised data is limited \cite{gui2024survey}.

As in other fields, recent work using neural networks to model symbolic music has started to adopt SSL \cite{wang2025notagen, qu2024mupt, zeng2021musicbert,  wu2023clamp}. However, the symbolic music data that these models are trained on is typically created manually, in a labor-intensive process. Acquiring it at the scale common for other modalities (e.g., text, images, audio) is challenging. Consequently, successful research often involves training from scratch on datasets such as Lakh and IMSLP \cite{raffel2016learning, IMSLP2024}, with research problems formulated around tasks that directly align with these datasets (e.g. multi-track symbolic music generation). 
This contrasts with other domains where substantial efforts have produced generalist models trained at an extreme scale, such as LLaMA and CLIP \cite{radford2021learning, touvron2023llama}, which provide strong foundations for research in data-limited settings \cite{zhou2023lima, kolesnikov2020big}. These constraints on symbolic music research become particularly clear when considering advancements in the neighboring area of audio modeling, where large-scale models including AudioGen and AudioLM \cite{kreuk2022audiogen, borsos2023audiolm}, alongside their underlying neural audio codecs \cite{defossez2022high, zeghidour2021soundstream}, have driven a broad range of advancements in music generation \cite{agostinelli2023musiclm, copet2023simple, borsos2023soundstorm}, and where SSL has been applied at scale to develop effective, general-purpose embedding models \cite{hubert, baevski2020wav2vec}.

\begin{figure}[t]
  \centering
  \includegraphics[width=\columnwidth]{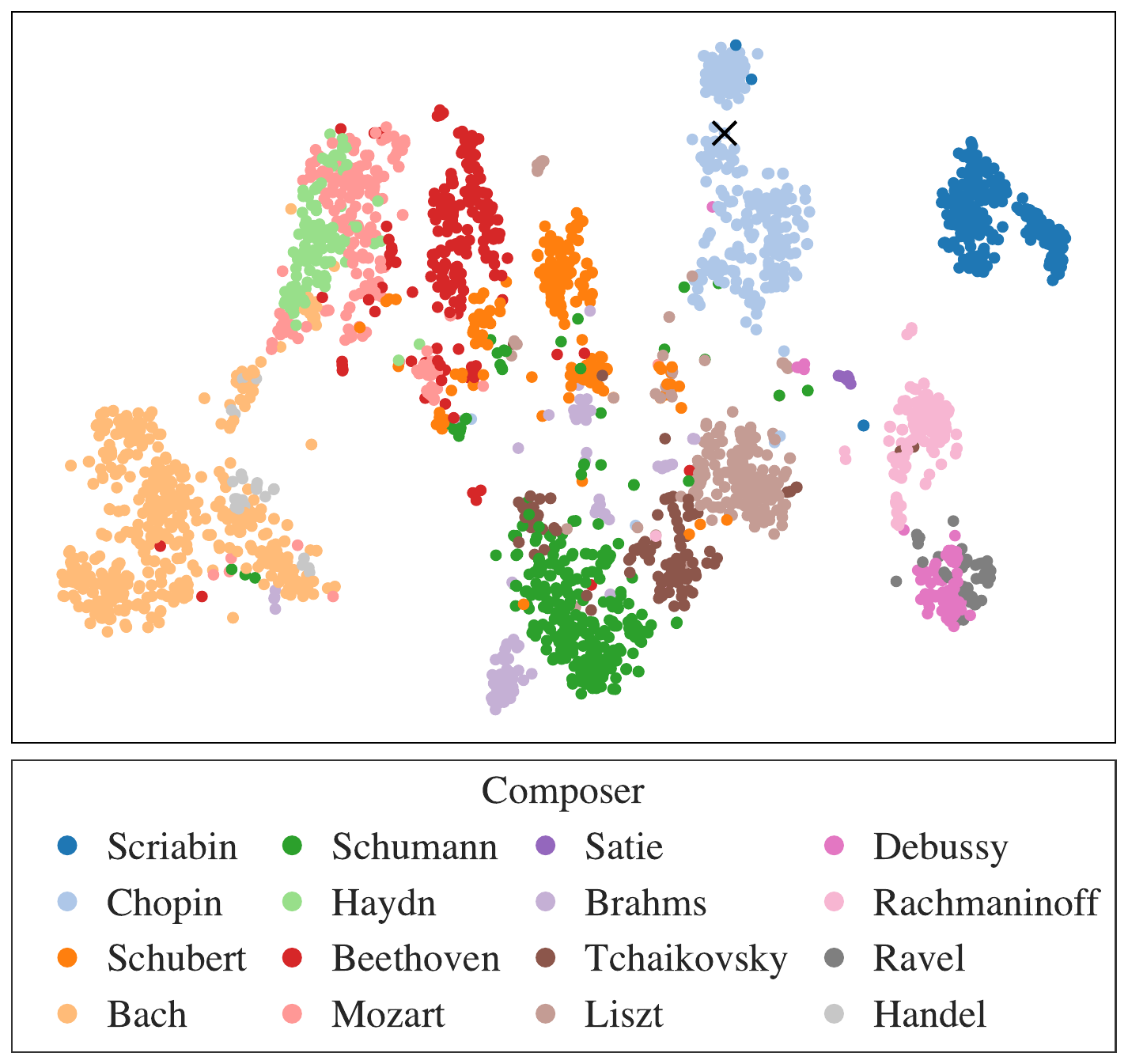}
  \caption
    [t-SNE visualisation of contrastive embeddings of classical compositions]
    {t-SNE visualisation of contrastive embeddings of classical compositions, trained on MIDI data without external metadata.  The cross (\textsf{×}) highlights Chopin’s \textit{Waltz in A minor}, which was discovered\footnotemark[1] after the training data was compiled, ensuring that
    it was not included.}
  \label{fig:tsne}
\end{figure}
\footnotetext[1]{See Javier C.~Hernández, “Hear a Chopin Waltz Unearthed After Nearly 200 Years,” \textit{The New York Times}, Oct.\,27, 2024.}
\addtocounter{footnote}{1}

Fortunately, strong progress has been made towards alleviating data bottlenecks for symbolic music research by leveraging neural networks trained for \textit{automatic music transcription} (AMT) \cite{benetos2018automatic}. In the restricted domain of solo-piano audio recordings, modern AMT models achieve highly reliable note-identification accuracy \cite{kong2021high, toyama2023automatic, yan2024scoring}, enabling automated dataset curation pipelines that crawl raw audio and transcribe it into MIDI using a combination of web scraping, audio-based processing, and AMT methods \cite{kong2020giantmidi, zhang2022atepp, edwards2023pijama}. Moreover, as this symbolic data is transcribed from real recordings, it captures the subtleties and dynamics of human performance. Recently, this combined progress has resulted in a new dataset of symbolic music, \textit{Aria-MIDI} \cite{bradshawaria}, comprising transcriptions of solo-piano recordings gathered at scale from YouTube, which has been made available for public use. At \textasciitilde100k hours, Aria-MIDI is orders of magnitude larger than similar datasets \cite{kong2020giantmidi}, presenting a unique opportunity to investigate the application of scaling SSL methods to symbolic music modeling.

Building on this, in this work we leverage Aria-MIDI to pretrain a generative transformer model via next-token prediction, using it as a foundation to explore the effectiveness of SSL techniques applied to symbolic music at a scale closer to recent applications in the text, image, and audio domains. We evaluate our model across two dimensions: generative modeling and representation learning. For generative capabilities, we conduct human listening tests comparing piano continuations generated by our model, while for representation learning we measure the ability of the pretrained model to adapt to MIR classification tasks via finetuning. To explore applications to similarity and retrieval tasks, we propose and analyze a novel self-supervised adaptation of the contrastive learning framework to symbolic music, which finetunes our model to produce embeddings that capture performance and composition-level features, as demonstrated by the natural composer clustering visualized in Figure~\ref{fig:tsne}. In both evaluation settings, we compare against symbolic and audio-based baselines. Overall, our experiments provide strong evidence that scaling SSL is a promising approach to tackling difficult tasks across symbolic MIR. Our key contributions are the following:

\begin{enumerate}
    \item We introduce and open-source \textit{Aria}\footnote{Available at: \href{https://github.com/eleutherai/aria}{https://github.com/eleutherai/aria}}, a pretrained autoregressive transformer model trained on transcriptions of piano recordings. Through human listening tests, we show it generates coherent continuations from short musical prompts, outperforming Anticipatory Music Transformer \cite{thickstun2023anticipatory} and rivaling proprietary audio models like Suno 3.5 \cite{suno}.
    
    \item We further demonstrate the effectiveness of large-scale pretrained representations for symbolic MIR through two approaches: (1) directly finetuning our model for classification tasks, achieving strong performance when labeled examples are extremely limited, and (2) proposing a novel adaptation of contrastive learning that produces an embedding model achieving state-of-the-art accuracy in linear probe experiments including composer, genre, and style detection. Critically, we show that this contrastive approach is effective \textit{only} when applied as a secondary finetuning phase.
\end{enumerate}

In addition to our models, we release a MIDI preprocessing and tokenization library designed to scale to large datasets and, although this work focuses on solo piano, to natively support multi-track MIDI files. Together, these contributions may serve as a foundation for future research in symbolic music modeling.

\section{Related Work}

Our work relates to many sub-areas of computational music, generative modeling, and representation learning. In this section, we focus on related work specific to the subfield of symbolic music modeling. 

The field of symbolic music generation using neural networks has advanced rapidly. Prior to the introduction of transformers, models such as DeepBach \cite{hadjeres2017deepbach} and Coconet \cite{huang2019counterpoint} demonstrated that neural networks are effective tools for modeling musical harmonies in Baroque music. The autoregressive paradigm for symbolic music generation, which models music as a stream of \textit{tokens}, gained traction by adapting architectures from natural language processing \cite{liang2017automatic}. This approach was extended by \cite{oore2020time} to incorporate expressive onset and duration timings, enabling generated music to more closely emulate human performance.

Music Transformer \cite{huang2018music} was a seminal work demonstrating the power and scalability of the autoregressive approach. The authors trained a transformer decoder on the MAESTRO dataset \cite{hawthorne2018enabling}, a collection of expressive MIDI piano recordings, and showed that autoregressive models could effectively learn long-term musical dependencies. Subsequent work from the same authors provided strong evidence that the musical and creative capabilities of their model scale well with dataset size \cite{simon2019pianotransformer}, reinforcing the value of curating large-scale piano transcription datasets as a future direction, a central premise we explore in our work.

Building on this foundation, MuseNet \cite{payne2019} expanded this approach by adding multi-track support to its MIDI tokenizer and training a larger model on a diverse corpus of multi-instrument data, including MAESTRO. Alternative tokenization schemes, such as REMI \cite{huang2020pop}, have also been influential. Variations of REMI have been adopted by models including Museformer \cite{yu2022museformer}, Figaro \cite{von2022figaro}, and MuseCoco \cite{lu2023musecoco}, which all introduced methods for conditioning generation on various musical features. Other research has explored representations beyond MIDI, such as the ABC notation \cite{walshaw2008abc} used by MuPT \cite{qu2024mupt}. More recently, Anticipatory Music Transformer \cite{thickstun2023anticipatory} was introduced as a versatile, state-of-the-art model for prompt continuation and infilling tasks with expressive millisecond-level precision.

For representation learning, several methods have been developed to produce symbolic music embeddings, useful as feature extractors for downstream classification tasks. These include MusicVAE \cite{roberts2018hierarchical}, a variational autoencoder for capturing long-term structure; MusicBERT \cite{zeng2021musicbert}, which learns self-supervised representations via a bar-masking objective; and the CLaMP series of models \cite{wu2023clamp, wu2024clamp, wu2025clamp}, which employ contrastive learning techniques to build cross-modal representations with natural language descriptions.

\begin{figure*}[t]
\centering
\begin{tikzpicture}[
    note/.style={rectangle, draw, fill=black, minimum width=2.25cm, minimum height=0.3cm},
    token/.style={rectangle, rounded corners, draw, fill=blue!20, minimum size=0.9cm, text width=1cm, text=black, align=center, font=\tiny}, 
    label/.style={text width=2.5cm, align=center, font=\small}, 
    pitch/.style={font=\small\bfseries},
    timeline/.style={thin}
]

\def\pianorolloffset{0.85}  % Change this value to offset horizontally

% Piano roll
\def\pianorollstart{0.5}
\def\pianorollend{10.5}
\draw[rectangle, draw=black, fill=gray!10] (\pianorollstart + \pianorolloffset, -1) rectangle (\pianorollend + \pianorolloffset, 1);
\node[label] at (-1, 0) {\textbf{PIANO-ROLL}};

% Timeline for pianoroll
\def\timestep{(\pianorollstart - \pianorollend - 1) / 8}
\draw[timeline] (\pianorollstart + 0.5 + \pianorolloffset, -0.75) -- (\pianorollend - 0.5 + \pianorolloffset, -0.75);
\draw[timeline] (\pianorollstart + 0.5 + \pianorolloffset, -0.8) -- (\pianorollstart + 0.5 + \pianorolloffset, -0.7);
\node[below, font=\tiny] at (\pianorollstart + 0.5 + \pianorolloffset, -0.7) {0};
\draw[timeline] (\pianorollstart + 0.5 + 1.125 + \pianorolloffset, -0.8) -- (\pianorollstart + 0.5 + 1.125 + \pianorolloffset, -0.7);
\node[below, font=\tiny] at (\pianorollstart + 0.5 + 1.125 + \pianorolloffset, -0.7) {1};
\draw[timeline] (\pianorollstart + 0.5 + 2.25 + \pianorolloffset, -0.8) -- (\pianorollstart + 0.5 + 2.25 + \pianorolloffset, -0.7);
\node[below, font=\tiny] at (\pianorollstart + 0.5 + 2.25 + \pianorolloffset, -0.7) {2};
\draw[timeline] (\pianorollstart + 0.5 + 3.375 + \pianorolloffset, -0.8) -- (\pianorollstart + 0.5 + 3.375 + \pianorolloffset, -0.7);
\node[below, font=\tiny] at (\pianorollstart + 0.5 + 3.375 + \pianorolloffset, -0.7) {3};
\draw[timeline] (\pianorollstart + 0.5 + 4.5 + \pianorolloffset, -0.8) -- (\pianorollstart + 0.5 + 4.5 + \pianorolloffset, -0.7);
\node[below, font=\tiny] at (\pianorollstart + 0.5 + 4.5 + \pianorolloffset, -0.7) {4};
\draw[timeline] (\pianorollstart + 0.5 + 5.625 + \pianorolloffset, -0.8) -- (\pianorollstart + 0.5 + 5.625 + \pianorolloffset, -0.7);
\node[below, font=\tiny] at (\pianorollstart + 0.5 + 5.625 + \pianorolloffset, -0.7) {5};
\draw[timeline] (\pianorollstart + 0.5 + 6.75 + \pianorolloffset, -0.8) -- (\pianorollstart + 0.5 + 6.75 + \pianorolloffset, -0.7);
\node[below, font=\tiny] at (\pianorollstart + 0.5 + 6.75 + \pianorolloffset, -0.7) {6};
\draw[timeline] (\pianorollstart + 0.5 + 7.875 + \pianorolloffset, -0.8) -- (\pianorollstart + 0.5 + 7.875 + \pianorolloffset, -0.7);
\node[below, font=\tiny] at (\pianorollstart + 0.5 + 7.875 + \pianorolloffset, -0.7) {7};
\draw[timeline] (\pianorollstart + 0.5 + 9.0 + \pianorolloffset, -0.8) -- (\pianorollstart + 0.5 + 9.0 + \pianorolloffset, -0.7);
\node[below, font=\tiny] at (\pianorollstart + 0.5 + 9.0 + \pianorolloffset, -0.7) {8};
\node[below, font=\tiny] at (\pianorollstart + 5.0 + \pianorolloffset, -1) {Time (seconds)};

\node[pitch] at (\pianorollstart + 1.125 + 0.25 + \pianorolloffset, -0.4) {C};
\filldraw[color=black, fill=blue!60] (\pianorollstart + 1.125 + 0.5 + \pianorolloffset, -0.55) rectangle ++ (3.375, 0.3);

\node[pitch] at (\pianorollstart + 3.375 + 0.25 + \pianorolloffset, 0.1) {E};
\filldraw[color=black, fill=blue!60] (\pianorollstart + 3.375 + 0.5 + \pianorolloffset, -0.05) rectangle ++ (3.375, 0.3);

\node[pitch] at (\pianorollstart + 5.625 + 0.25 + \pianorolloffset, 0.6) {G};
\filldraw[color=black, fill=blue!60] (\pianorollstart + 5.625 + 0.5 + \pianorolloffset, 0.45) rectangle ++ (3.375, 0.3);

% Tokenization 1
\node[label] at (-1, -2) {\textbf{MUSIC\\TRANSFORMER}};
\node[token] at (\pianorollstart + 0.57, -2) {\texttt{SHIFT\\1000MS}};
\node[token, fill=yellow!30] at (\pianorollstart + 1.90, -2) {\texttt{SET\_VEL\\60}};
\node[token, fill=green!20] at (\pianorollstart + 3.22, -2) {\texttt{NOTE\_ON\\60}};
\node[token] at (\pianorollstart + 4.54, -2) {\texttt{SHIFT\\1000MS}};
\node[token] at (\pianorollstart + 5.87, -2) {\texttt{SHIFT\\1000MS}};
\node[token, fill=green!20] at (\pianorollstart + 7.19, -2) {\texttt{NOTE\_ON\\64}};
\node[token] at (\pianorollstart + 8.51, -2) {\texttt{SHIFT\\1000MS}};
\node[token, fill=red!20] at (\pianorollstart + 9.83, -2) {\texttt{NOTE\_OFF\\60}};
\node[token] at (\pianorollstart + 11.16, -2) {\texttt{SHIFT\\1000MS}};
\node at (\pianorollstart + 12.31, -2) {\ldots};

% Tokenization 2
\node[label] at (-1, -3) {\textbf{MUSENET}};
\node[token] at (\pianorollstart + 0.57, -3) {\texttt{WAIT\\1000MS}};
\node[token, fill=green!20] at (\pianorollstart + 1.90, -3) {\texttt{PIANO\\P: C4\\V: 60}};
\node[token] at (\pianorollstart + 3.22, -3) {\texttt{WAIT\\2000MS}};
\node[token, fill=green!20] at (\pianorollstart + 4.54, -3) {\texttt{PIANO\\P: E4\\V: 60}};
\node[token] at (\pianorollstart + 5.87, -3) {\texttt{WAIT\\1000MS}};
\node[token, fill=red!20] at (\pianorollstart + 7.19, -3) {\texttt{PIANO\\P: C4\\V: 0}};
\node[token] at (\pianorollstart + 8.51, -3) {\texttt{WAIT\\1000MS}};
\node[token, fill=green!20] at (\pianorollstart + 9.83, -3) {\texttt{PIANO\\P: G4\\V: 60}};
\node[token] at (\pianorollstart + 11.16, -3) {\texttt{WAIT\\1000MS}};
\node at (\pianorollstart + 12.31, -3) {\ldots};

% Tokenization 3
\node[label] at (-1, -4) {\textbf{REMI}};
\node[token, fill=yellow!30] at (\pianorollstart + 0.57, -4) {\texttt{BAR}};
\node[token] at (\pianorollstart + 1.90, -4) {\texttt{POSITION\\2/4}};
\node[token, fill=green!20] at (\pianorollstart + 3.22, -4) {\texttt{PITCH\\C4}};
\node[token, fill=green!20] at (\pianorollstart + 4.54, -4) {\texttt{VELOCITY\\60}};
\node[token, fill=red!20] at (\pianorollstart + 5.87, -4) {\texttt{DURATION\\3/4}};
\node[token, fill=blue!20] at (\pianorollstart + 7.19, -4) {\texttt{POSITION\\4/4}};
\node[token, fill=green!20] at (\pianorollstart + 8.51, -4) {\texttt{PITCH\\E4}};
\node[token, fill=green!20] at (\pianorollstart + 9.83, -4) {\texttt{VELOCITY\\60}};
\node[token, fill=red!20] at (\pianorollstart + 11.16, -4) {\texttt{DURATION\\3/4}};
\node at (\pianorollstart + 12.31, -4) {\ldots};

% ARIA 
\node[label] at (-1, -5) {\textbf{ARIA}};
\node[token, fill=green!20] at (\pianorollstart + 0.57, -5) {\texttt{PIANO\\P: 60\\V: 60}};
\node[token] at (\pianorollstart + 1.90, -5) {\texttt{ONSET\\1000MS}};
\node[token, fill=red!20] at (\pianorollstart + 3.22, -5) {\texttt{DURATION\\3000MS}};
\node[token, fill=green!20] at (\pianorollstart + 4.54, -5) {\texttt{PIANO\\P: 64\\V: 60}};
\node[token] at (\pianorollstart + 5.87, -5) {\texttt{ONSET\\3000MS}};
\node[token, fill=red!20] at (\pianorollstart + 7.19, -5) {\texttt{DURATION\\3000MS}};
\node[token, fill=yellow!30] at (\pianorollstart + 8.51, -5) {\texttt{<T>}};
\node[token, fill=green!20] at (\pianorollstart + 9.83, -5) {\texttt{PIANO\\P: 67\\V: 60}};
\node[token] at (\pianorollstart + 11.16, -5) {\texttt{ONSET\\0MS}};
\node at (\pianorollstart + 12.31, -5) {\ldots};

\end{tikzpicture}
\vspace{0.3cm}
\caption{Comparison of different tokenizations of a piano-roll, using various approaches. Music Transformer \cite{huang2018music} and MuseNet \cite{payne2019} track the passage of time using time-shift tokens, whereas Aria uses absolute onsets relative to the current segment. The REMI tokenizer \cite{huang2020pop} uses a neural beat-tracking model to estimate positions of notes and bar delimiters \cite{bock2016joint}.}
\label{fig:tokens}
\end{figure*}
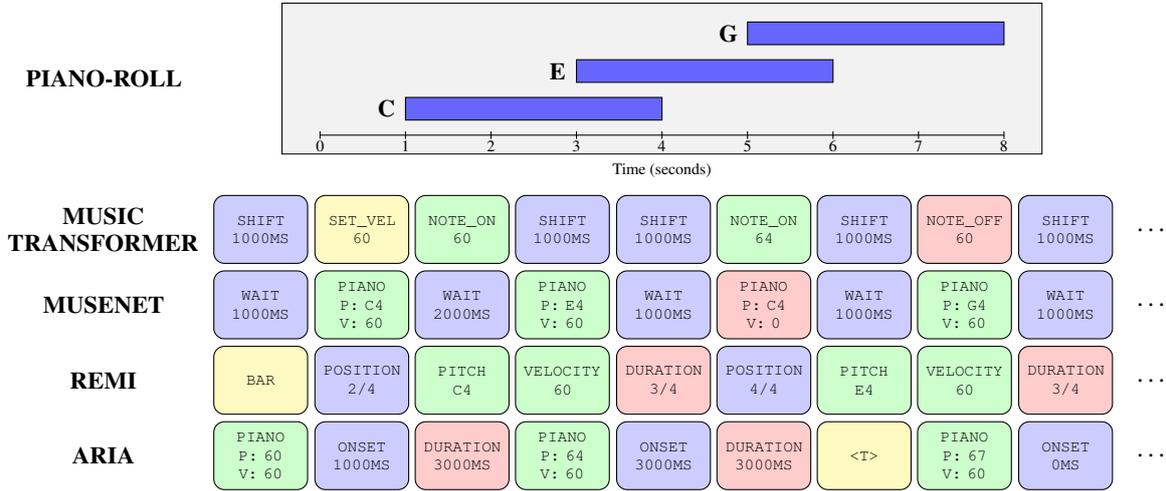

\section{Method}
\label{sec:method}

To explore the capabilities of large-scale self-supervised models for piano performance, we first pretrained an autoregressive transformer model using next-token prediction on a refined subset of the Aria-MIDI dataset. We adopt this setup due to its versatility: next-token prediction has a proven track record in generative modeling for both symbolic and audio-based music \cite{huang2018music, borsos2023audiolm}, as well as adaptability to downstream tasks via finetuning \cite{raffel2020exploring}. Apart from the tokenization scheme, which we hand-designed, we used a conventional modern transformer architecture with minimal modifications, providing a standardized foundation for evaluating our hypothesis and supporting further research.

\subsection{MIDI Tokenization}
\label{sec:token}

To autoregressively model MIDI files as streams of discrete tokens, we chose to use a temporal resolution of 10 milliseconds for note onsets and durations, and discretize note velocity values into 12 bins. Our tokenizer is designed to natively handle multi-track (multi-instrument) MIDI files by condensing the 128 MIDI instruments, corresponding to \textttalt{program\_change} MIDI messages, into 13 instrument classes, including one for percussion.

Given a MIDI file, we resolve its constituent \textttalt{note\_on} and \textttalt{note\_off} events into a list of notes. For non-percussion instruments, we tokenize a note with pitch $p$, velocity $v$, and absolute onset/offset in milliseconds $(t_{\text{on}}, t_{\text{off}})$ as a triple of tokens:
$$ [\text{instrument}, p, v], [\text{onset:}\: t_{\text{on}}], [\text{duration:}\: t_{\text{off}} -t_{\text{on}} ] $$
For percussion, we tokenize a note with note number $n$ and onset $t_{\text{on}}$ as:
$$ [\text{drum}, n], [\text{onset:}\: t_{\text{on}}] $$
The tokenization of an entire MIDI file is constructed by concatenating the tokenizations of the constituent notes in order of onset. MIDI metadata, such as key, tempo, and time signature, is discarded, and other relevant musical information, such as the sustain pedal, is incorporated directly into the duration tokens. This schema is set apart from some popular tokenization techniques used for symbolic music, such as REMI \cite{huang2020pop} and text-based score representations ABC \cite{walshaw2008abc} and MusicXML \cite{good2001musicxml}, as it does not include beat or bar information, instead representing onsets and durations in milliseconds.

In the MIDI standard \cite{midi1996}, \textttalt{note\_on} and \textttalt{note\_off} events are spaced temporally by specifying a number of ticks to wait before processing the next event. For Music Transformer and MuseNet, the authors incorporate this into their chosen MIDI tokenization schemes \cite{huang2018music, payne2019}, using \textit{time-shift} tokens to separate notes rather than specifying their absolute onset times.  However, emerging work has provided evidence that using time-shift tokens in this way may be suboptimal in transformer-based models, resulting in reduced accuracy in sequence-to-sequence piano transcription\cite{hawthorne2021sequence}, and unstable rhythm or drifting bar lines in musical generations \cite{huang2020pop}. One possible explanation is that when using \textit{relative-timing} tokenization, autoregressive models struggle to maintain an exact temporal representation of the prior context, as they must sum up many sequential time-shift values to calculate temporal relationships between notes with medium or long-term dependencies. Previous studies on large language models have demonstrated that transformers can struggle with exactly this sort of arithmetic \cite{lee2023teachingarithmeticsmalltransformers, mcleish2024transformersarithmeticrightembeddings}.

In preliminary investigations, we also observed negative effects when using relative-timing tokenizations, particularly on temporal instability in passages with rapid note sequences. To address these issues, we chose to adopt \textit{absolute onset times} in our tokenizer. We implemented this by dividing the music into 5000-millisecond segments and recording note onsets relative to the start of each segment -- this helped us avoid expanding the tokenizer’s vocabulary to include all possible absolute onset times. To remove ambiguity, we marked the start of each new segment using a special token: \texttt{<T>}.  We designed this to resemble note timing using beat-position within a bar, however, unlike tokenization schemes that do this directly \cite{huang2020pop, hsiao2021compound}, our approach is applicable to MIDI files that lack beat and bar information, such as those transcribed from solo piano recordings. Figure~\ref{fig:tokens} illustrates how our approach differs from other approaches.
\begin{equation}
\label{eq:time}
  T_i - T_j = 
  \begin{cases}
    \textstyle\sum_{k=j+1}^{i} w_k & \text{Relative} \\[1ex]
    C(\texttt{<T>}, i, j) + \tilde{o}_i - \tilde{o}_j & \text{Hybrid (Ours)} \\[1ex]
    o_i - o_j & \text{Absolute}
  \end{cases}
\end{equation}
Equation~\ref{eq:time} demonstrates the arithmetic required to calculate the time separating two notes $n_i$ and $n_j$ across the different tokenization approaches, where $w_k$ denotes the length of the time-shift message preceding note $k$, $C(\texttt{<T>}, i, j)$ represents the total time spanned by complete 5000ms segments between notes $n_i$ and $n_j$, calculated by counting the number of segment tokens and multiplying by the segment duration, $o_k$ represents the absolute onset time of note $k$, and $\tilde{o}_k$ represents the adjusted absolute onset time of note $k$ relative to the start of its 5000ms segment.

\subsection{Model}
\label{sec:model}

Our model architecture builds upon the LLaMa 3.2 model family, chosen due to its effectiveness in autoregressive tasks across modalities \cite{grattafiori2024llama}. Using the 1B parameter configuration as a starting point, we made several architectural modifications. Firstly, guided by established principles on model-data ratios for language models \cite{hoffmann2022training}, we reduced the hidden state dimension ($\text{d}_{\text{model}}$) from 2048 to 1536. This decreased the parameter count by roughly half, balancing model capacity with computational efficiency for our dataset scale. Secondly, we simplified the architecture by opting for standard multi-head attention (with 24 heads) and layer normalization \cite{vaswani2017attention, ba2016layer}, instead of grouped-query attention and RMS normalization as used in standard LLaMa 3 variants \cite{ainslie2023gqa, zhang2019root}.

\textbf{Pretraining dataset.} As our training corpus consists of automatically transcribed internet-sourced piano recordings, significant variability exists in transcription quality and content suitability, potentially introducing harmful biases or noisy data into downstream models. To mitigate this, we implemented rigorous preprocessing steps. To reduce memorization, we addressed extreme cases of composition duplication, such as repeated performances of overrepresented works, by applying filtering based on compositional metadata. Specifically, for composers with more than 250 instances of files containing opus and/or piece number tags, we retained at most 10 instances per opus/piece-number pair. For these same composers, we also discarded all other files that lack compositional identifiers. Additionally, we employed heuristic-based filtering, considering note density, pitch and duration entropy, silence, and indicators of repetitive content, to exclude problematic entries (e.g., \textit{Black MIDI}\footnote{\href{https://en.wikipedia.org/wiki/Black\_MIDI}{https://en.wikipedia.org/wiki/Black\_MIDI}}). Following these steps, our refined pretraining corpus comprises 820,944 MIDI files, amounting to 60,473 hours of solo piano music.

\textbf{Pretraining recipe.} We pretrained our model using standard next-token prediction on concatenated sequences of tokenized MIDI files, as detailed in Section~\ref{sec:token}. A sequence length of 8192 tokens was chosen to balance computational constraints with the need to learn meaningful short- and long-term dependencies within piano music. To enhance generalization and prevent overfitting, we utilized online data augmentation, randomly transposing ($\pm$5 semitones), varying tempo ($\pm$20\%), and adjusting MIDI velocity ($\pm$10). 

\textbf{Generative finetuning.} We produced a model variant tailored for generative piano-continuation tasks by applying a single-epoch finetuning phase after pretraining, annealing the learning rate to zero while training on higher-quality data. To enhance data quality, we removed all identified compositional duplicates, tightened existing quality filters, and introduced an additional filter aimed at excluding transcriptions of synthesized MIDI files\footnote{Preprocessing details: \href{https://github.com/loubbrad/aria-midi}{https://github.com/loubbrad/aria-midi}}\hspace{-1mm}. Additionally, during this phase, each training sequence begins at the start of a new file (i.e., non-concatenated), and we insert a special token (\texttt{<D>}) approximately 100 tokens before the end of each training example to enable explicit inference-time control over generation endings.

\subsection{Contrastive Representation Learning}
\label{sec:method_con}

To investigate the strength of the pretrained representations, we propose a secondary finetuning stage, adapting the pretrained model to generate \textit{embeddings} of tokenized sequences. Our approach leverages the SimCLR framework for contrastive representation learning \cite{chen2020simple}. In SimCLR, an encoder is trained to produce similar embeddings for different \textit{views} of the same training example while simultaneously pushing embeddings from unrelated examples apart through minimization of a contrastive loss. This approach has demonstrated strong results in music, capturing semantic relationships within embeddings effectively \cite{spijkervet2021contrastive, choi2022towards}, and has recently been combined with large pretrained language models to produce rich textual embeddings \cite{gao2021simcse, wang2023improving}.

To generate two distinct views of a MIDI file, we randomly extract two different contiguous slices, each comprising between 100 and 650 notes (approximately 300–2000 tokens). Each slice undergoes independent data augmentation using our standard procedures before tokenization. To produce sequence embeddings, we replace the original language modeling head with an embedding head, projecting the final hidden state into a 512-dimensional embedding space. We derive a slice’s embedding from the hidden state associated with an end-of-sequence token appended after the final note token-triple. 

To calculate the contrastive loss, we use the normalized temperature-scaled cross-entropy loss, \textit{NT-Xent}, over minibatches of related embedding pairs:
\begin{equation}
    \label{eq:loss_fn}
    \ell_{i, j}=-\log \frac{\exp \left(\operatorname{sim}\left(z_{i}, z_{j}\right) / \tau\right)}{\sum_{k=1}^{2 N} \mathbbm{1}_{[k \neq i]} \exp \left(\operatorname{sim}\left(z_{i}, z_{k}\right) / \tau\right)}
\end{equation}
Here, $\operatorname{sim}(z_k, z_l)$ denotes the cosine similarity between normalized embeddings $z_k$ and $z_l$, $\mathbbm{1}_{[k \neq i]} \in \{0, 1\}$ is an indicator function, and $\tau$ is the temperature parameter. Each minibatch consists of $N$ MIDI files, from which we construct $N$ pairs of related embeddings (i.e., $2N$ total embeddings), $\{z_i, z_{i+N}\}_{i=1,\dots,N}$, where both $z_i$ and $z_{i+N}$ are derived from two augmented views of the same file. We train the model by minimizing the symmetric loss: $L :=  \frac{1}{2}\sum_{k=1}^N (\ell_{k, k+N} + \ell_{k+N, k})$.

This setup has two key advantages. First, by extracting non-overlapping slices from the same file, the model learns embeddings reflecting higher-level musical semantics such as genre, composer, style, and performance nuances, rather than local details. This is important for musical performances, where standard supervised representation learning approaches, e.g., MuLan \cite{huang2022mulan}, are limited due to the descriptive subtlety and complexity of musical attributes. Second, our approach facilitates studying how effectively next-token prediction representations transfer to contrastive embedding frameworks. When trained from scratch, SimCLR-inspired training methods typically require large amounts of in-batch negatives, which pose significant VRAM constraints \cite{chen2020simple}. However, recent work on text embeddings shows that initializing contrastive training from pretrained models can alleviate this \cite{gao2021simcse}. Thus, our method introduces a general-purpose semi-supervised framework for representation learning of symbolic music, which allows us to evaluate the transferability of next-token musical representations.

\section{Experiments}
\label{sec:setup}

Having outlined our methodology, we evaluate the generative capabilities of our model, as well as the contrastive representation learning framework, in the context of piano performance. To understand its capabilities in the wider area of models for generative music and MIR, we compare our approach to both symbolic and audio-based baselines, utilizing Pianoteq \cite{pianoteq} to synthesize MIDI files into audio.

\subsection{Setup}

We pretrained our model using the AdamW optimizer for 75 epochs over the training corpus. We used a learning rate of $3e\text{-}4$ with 1000 warmup steps, followed by a linear decay to 10\% of the initial rate over the course of training. The model has approximately 650 million parameters and was pretrained for 9 days on 8 H100 GPUs with a batch size of 16 per GPU. 

In the contrastive finetuning stage, we used a learning rate of $1e\text{-}5$ with the same linear decay schedule. We set the NT-Xent temperature parameter to $\tau = 0.1$. This phase lasted 25 epochs, during which each MIDI file contributes exactly one pair of augmented views per epoch. We trained on the reduced finetuning dataset described in Section~\ref{sec:model}; however, we relaxed the preprocessing constraints on compositional duplicates to encourage the model to distinguish between different performances of popular compositions.

\begin{table}[t]
\centering
\renewcommand{\arraystretch}{1.25}
\setlength{\tabcolsep}{6pt}
\fontsize{9.7}{8}\selectfont
\begin{tabular}{l c c c c}
\toprule
\textbf{Compared Model} & \textbf{Wins} & \textbf{Ties} & \textbf{Losses} & \textbf{p-value} \\
\midrule
AM Transformer & $38$ & $0$ & $6$ & $9.43e\text{-}7$ \\
Suno 3.5       & $18$ & $9$ & $21$ & $7.49e\text{-}1$ \\
MusicGen       & $49$ & $1$ & $0$ & $3.55e\text{-}15$ \\
Ground Truth   & $15$ & $9$ & $17$ & $8.60e\text{-}1$ \\
\bottomrule
\end{tabular}
\caption{Pairwise human preference results comparing musical coherence of 45-second continuations of 15-second prompts. We report the number of times our model won, tied, or lost against the listed model. P-values are computed using a two-sided binomial test on non-tied comparisons.}
\label{tab:human_eval}
\end{table}

\textbf{Generative modeling.} Following the generative finetuning procedure described in Section~\ref{sec:model}, we explore the generative capabilities of the resulting model by analyzing the \textit{musical coherence} of continuations of short solo piano prompts. This methodology aligns with evaluations in previous work \cite{borsos2023audiolm, thickstun2023anticipatory}, and mitigates taste bias by having participants evaluate continuations within the same musical style.

In our listening test, we asked 46 participants with at least one year of musical training to compare 45-second continuations generated from 15-second solo piano prompts, evaluating their musical coherence. Participants were presented with a series of random pairwise A/B comparisons, where they were asked to indicate their preferred continuation, guided by criteria such as melodic development, rhythmic structure, harmonic progression, and stylistic coherence. To generate test samples, we selected five prompts representing different subgenres of solo piano music, and generated eight continuations per prompt (totaling 40 continuations per model). We compared our model’s outputs against several baselines, including Anticipatory Music Transformer (\textttalt{music-large-800k}) \cite{thickstun2023anticipatory}, the audio-based generative models MusicGen (\textttalt{large}) \cite{agostinelli2023musiclm} and Suno 3.5 \cite{suno}, and the human-composed ground-truth.

\textbf{Contrastive embeddings.} We evaluate our approach to learning contrastive embeddings by training linear classifiers on the frozen embeddings produced by different models and comparing their performance on held-out test sets. We assess performance using established benchmarks, Pianist8 \cite{pianist8} and VG-MIDI \cite{vgmidi}, as well as new benchmarks we derive from Aria-MIDI metadata. Specifically, we extracted label-balanced train-test splits comprising 10,000 and 1,000 files, respectively, for four classification tasks: Genre (2 classes), Musical Period (4 classes), Form (6 classes), and Composer (10 classes). For comparison, we include results from CLaMP 3 (\textttalt{saas}) \cite{wu2025clamp}, M3 \cite{wu2024clamp}, and the audio-based model MERT \cite{li2023mert}. Linear classifiers were trained on global file embeddings obtained by averaging slice embeddings within each file. We trained with a learning rate of $3e\text{-}4$ and a linear decay schedule to 0, running separate experiments with 10, 20, and 50 epochs, and reporting the best result.

\begin{table*}[t]
\centering
\renewcommand{\arraystretch}{1.25}
\setlength{\tabcolsep}{6pt} 
\fontsize{9.7}{8}\selectfont
\begin{tabular}{l c c c c c c c c c c c c}
\toprule
\multirow{2}{*}{\textbf{Model}} 
  & \multicolumn{2}{c}{\textbf{Genre}} 
  & \multicolumn{2}{c}{\textbf{Form}} 
  & \multicolumn{2}{c}{\textbf{Musical Period}} 
  & \multicolumn{2}{c}{\textbf{Composer}} 
  & \multicolumn{2}{c}{\textbf{Pianist8}}
  & \multicolumn{2}{c}{\textbf{VG-MIDI}} \\
\cmidrule(lr){2-3} \cmidrule(lr){4-5} \cmidrule(lr){6-7} \cmidrule(lr){8-9} \cmidrule(lr){10-11} \cmidrule(lr){12-13}
  & \textbf{Acc} & \textbf{F1}
  & \textbf{Acc} & \textbf{F1}
  & \textbf{Acc} & \textbf{F1}
  & \textbf{Acc} & \textbf{F1}
  & \textbf{Acc} & \textbf{F1}
  & \textbf{Acc} & \textbf{F1} \\
\midrule
\textit{Main Results} & & & & & & & & & & & & \\
\addlinespace[0.5ex]
MERT                & 83.00 & 83.00 & 63.89 & 63.90 & 69.50 & 68.94 & 69.60 & 69.30 & 65.06 & 65.18 & 45.45 & 40.37 \\
M3                  & 85.10 & 85.10 & 69.88 & 70.12 & 71.20 & 70.81 & 71.90 & 71.72 & 81.93 & 81.48 & 54.55 & 46.13 \\
CLaMP 3             & 89.10 & 89.10 & 77.79 & 77.97 & 80.60 & 80.20 & 84.50 & 84.46 & 80.72 & 79.76 & 45.45 & 36.53 \\
Aria$_{\text{Emb}}$ & \underline{92.40} & \underline{92.40} & \underline{82.45} & \underline{82.57} & \underline{84.70} & \underline{84.69} & \underline{90.50} & \underline{90.49} & \textbf{91.57} & \textbf{92.38} & \underline{63.64} & \underline{63.96} \\
Aria$_{\text{Ft}}$  & \textbf{93.20} & \textbf{93.20} & \textbf{87.53} & \textbf{87.59} & \textbf{86.50} & \textbf{86.53} & \textbf{96.30} & \textbf{96.32} & \underline{91.56} & \underline{92.03} & \textbf{68.18} & \textbf{69.55} \\
\midrule
\textit{Embeddings} & & & & & & & & & & & & \\
\addlinespace[0.5ex]
Aria$_{e=25}^{\dag}$ & 82.30 & 82.30 & 66.94 & 66.96 & 69.00 & 68.50 & 65.50 & 65.41 & 84.34 & 84.56 & 59.09 & 54.29 \\
Aria$_{e=1}$            & 92.90 & 92.90 & 80.53 & 80.69 & 83.80 & 83.71 & 87.60 & 87.62 & 92.77 & 93.71 & 59.09 & 57.80 \\
Aria$_{\tau=0.05}$      & 92.40 & 92.40 & 81.34 & 81.48 & 84.00 & 83.85 & 89.90 & 89.90 & 95.18 & 95.71 & 59.09 & 54.32 \\
Aria$_{\tau=0.5}$       & 92.30 & 92.30 & 73.43 & 73.63 & 80.70 & 80.56 & 70.20 & 70.05 & 91.57 & 92.70 & 54.55 & 45.00 \\
\midrule
\textit{Finetuning} & & & & & & & & & & & & \\
\addlinespace[0.5ex]
Aria$_{n=100}$  & 89.50 & 89.50 & 68.26 & 68.20 & 70.20 & 70.64 & 65.30 & 64.10 & -     & -     & - & - \\
Aria$_{n=200}$  & 91.10 & 91.10 & 75.25 & 75.54 & 75.10 & 75.68 & 78.10 & 78.08 & -     & -     & - & - \\
Aria$_{n=500}$  & 90.80 & 90.80 & 79.31 & 79.49 & 80.90 & 80.91 & 85.20 & 85.18 & -     & -     & - & - \\
Aria$_{n=1000}$ & 91.40 & 91.40 & 80.63 & 80.68 & 82.90 & 83.01 & 90.10 & 90.12 & -     & -     & - & - \\
\bottomrule
\end{tabular}
\caption{Classification performance across symbolic music tasks. We report maximum accuracy (Acc) and macro-F1 scores (F1) for each task. \textit{Main Results} compare our embedding model (Aria$_\text{Emb}$) and supervised finetuned model (Aria$_\text{Ft}$) to other models (MERT, M3, CLaMP 3). \textit{Embedding} ablations vary key components of the contrastive learning setup: training epochs ($e$), temperature parameter ($\tau$), and without pretraining (\dag), while keeping all other settings the same as Aria$_\text{Emb}$. \textit{Finetuning} ablations show test-set performance as a function of the number of labeled training files ($n$).}
\label{tab:classification_results}
\end{table*}

\textbf{Supervised finetuning.} To complement our linear probe experiments, we evaluate how well our pretrained model adapts to supervised musical classification tasks, employing finetuning techniques inspired by NLP literature \cite{devlin2019bert, raffel2020exploring}. For classifier finetuning, we replaced the language modeling head with a classification head, predicting labels directly from the hidden state of the end-of-sequence token. During this phase, we finetuned all model weights end-to-end using a learning rate of $1e\text{-}5$ (without warmup) with linear decay schedule, and applied dropout to residual connections, increasing the dropout rate linearly from $p_d = 0.0$ (first layer) to $p_d = 0.2$ (final layer). By systematically varying the number of labeled training examples, using class-balanced subsets, we analyze our pretrained model's ability to adapt to supervised symbolic MIR tasks in scenarios with limited labeled data. In each case, we trained for 10 epochs and report the results from the best-performing epoch.

\subsection{Results}

Table~\ref{tab:human_eval} reports the results of our listening test. Participants consistently preferred the musical coherence of continuations produced by our model over those from Anticipatory Music Transformer and MusicGen. This signals a notable improvement in symbolic models for piano performance generation, which we primarily attribute to the scale of our training dataset, given our standardized setup. It also highlights limitations in audio models like MusicGen, whose restricted context window necessitates sliding-window inference, diminishing coherence in longer generations. Conversely, we found no statistically significant preference difference between our model's outputs and either Suno 3.5 or human-composed ground-truth continuations. We acknowledge two key limitations: Firstly, we could not include closed-access models like AudioLM \cite{borsos2023audiolm}, despite their promising reported results on similar piano-continuation benchmarks. Secondly, our evaluation excludes popular symbolic models such as MuPT \cite{qu2024mupt}, as their bar-level timing representation (e.g., ABC notation) is incompatible with expressive millisecond-level MIDI performances.

Table~\ref{tab:classification_results} summarizes the results of our linear probe and supervised finetuning classification experiments, alongside an ablation study of training configurations for contrastive learning. Our proposed method for semi-supervised representation learning substantially improves results on all benchmarks, producing embeddings that capture diverse file-level musical attributes without incorporating metadata during training.  The ablation study further highlights the importance of initializing contrastive training from pretrained next-token representations, demonstrating that our contrastive method is competitive only when applied as a finetuning stage. Notably, finetuning on one embedding pair per file for a single epoch (Aria$_{e=1}$) surpasses training from scratch on 25 pairs per file (Aria$_{e=25}^{\dagger}$). While this represents an advancement, we note that our benchmarks focus exclusively on piano performances, whereas the comparison models support multi-instrument MIDI or audio files. Finally, our supervised finetuning experiments demonstrate the strong adaptability of next-token prediction SSL frameworks to supervised symbolic MIR tasks. Our finetuned models achieve state-of-the-art classification performance on large datasets and perform surprisingly well on complex tasks, even when trained on limited labeled data. 

\section{Conclusion}

We introduce Aria, an autoregressive generative transformer model designed to investigate the scalability of self-supervised learning for symbolic music modeling. Our experiments show that this pretraining framework effectively adapts to generative modeling, MIDI-embedding generation, and supervised MIR tasks. Moreover, our findings suggest that careful data curation and large-scale training can unlock new opportunities for downstream symbolic music applications, particularly in settings where data is scarce.

\section{Acknowledgments}

% TODO: Finalize
This work was supported by UKRI and EPSRC under grant EP/S022694/1. Additional support was provided by EleutherAI and StabilityAI, as well as a compute grant from the Ministry of Science and ICT of Korea and Gwangju Metropolitan City.

\bibliography{ISMIRtemplate}

\end{document}